# Understanding Super-Earths with MINERVA-Australis at USQ's Mount Kent Observatory

Robert A Wittenmyer[1], Jonathan Horner[1], Brad D Carter[1], Stephen R Kane[2], Peter Plavchan[3], David Ciardi[4] and the MINERVA-Australis consortium

[1] *University of Southern Queensland, Toowoomba, Qld 4350, Australia*
[2] *Department of Earth Sciences, University of California, Riverside, CA 92521, USA*
[3] *George Mason University, 4400 University Drive, Fairfax, VA 22030, USA*
[4] *NASA Exoplanet Science Institute, California Institute of Technology, 1200 East California Avenue, Pasadena, CA 91125, USA*

**Summary:** Super Earths, planets between 5-10 Earth masses, are the most common type of exoplanet known, yet are completely absent from our Solar system. As a result, their detailed properties, compositions, and formation mechanisms are poorly understood. NASA's Transiting Exoplanet Survey Satellite (TESS) will identify hundreds of Super-Earths orbiting bright stars, for the first time allowing in-depth characterisation of these planets.

At the University of Southern Queensland, we are host to the MINERVA-Australis project, dedicated wholly to the follow-up characterisation and mass measurement of TESS planets. We give an update on the status of MINERVA-Australis and our expected performance.

We also present results from the fully operational Northern MINERVA array, with the primary mission of discovering rocky planets orbiting 80 nearby bright stars.

## Introduction

The past two decades have seen the dawn of the Exoplanet Era – a revolution in our understanding of the universe around us. Prior to the mid-1990s, we wondered whether the Solar system was unique, or if planets were common around other stars. The ubiquity of planets was one of the key bones of contention in discussions of our own place in the universe. To systematically answer the question 'are we alone?', we must first understand whether planetary systems like our own are the exception or the rule.

The first steps in unravelling that mystery came with the discovery of debris disks, orbiting nearby stars, by the Infra-Red Astronomical Satellite (IRAS) in the 1980s [1][2][3]. That discovery offered a tantalising hint to the ubiquity of planetary systems, revealing that the processes thought to result in the formation of planets were not uncommon in the cosmos.

The discovery of the first planets orbiting other stars was announced in 1992, and was the first of many great surprises of the Exoplanet Era. Rather than being planets like our own, orbiting a Sun-like star, these first planets were instead small, battered husks, orbiting a pulsar – the highly compressed remains of a star that once went supernova [4].

The first planets found around Sun-like stars came soon after, with the discovery of Dimidium (51 Pegasi b; [5]), Taphao Thong (47 Ursa Majoris b; [6]) and 70 Virginis b ([7]) in 1995 and 1996. Following these discoveries, it was realised that objects found in the late 1980s (HD 114762 b; [8], and Gamma Cephei b; [9]) were also planetary in nature, though they had been considered brown dwarfs at the time of their discovery. These first planets were drastically different to those seen in the Solar system. Where our planetary system contains small, rocky worlds (the telluric planets) close to the Sun, and giant planets further out, the first discovered exoplanets were instead giant planets on relatively short period orbits.

During the first decade of the Exoplanet Era, the great majority of new planets found were discovered using the radial velocity technique (e.g. [10][11][12]). That method uses observations of the lines in a star's spectrum to follow the subtle backward and forward wobbles the star performs as a result of the gravitational pull of a planet. The more massive the planet, and the closer that planet to the star, the larger the wobble – which makes the radial velocity method strongly biased towards the detection of massive planets on relatively short period orbits. Despite these drawbacks, the increasing timelines over which radial velocity observations have been carried out have more recently facilitated the discovery of the first true Jupiter analogues – planets whose mass and orbits more closely resemble those of the gas giants in the Solar system (e.g. [13][14][15]).

More recently, the rapid explosion in the number of planets found using the transit technique has overshadowed the success of the radial velocity method. The first transiting planets were found early in the 21$^{st}$ century ([16][17][18]). These early discoveries hinted to the great promise of the technique, which is facilitated by the exquisite photometric precision afforded to observers by the latest generations of astronomical imagers.

In 2009, NASA launched the *Kepler* observatory, which proceeded to spend some four years staring continually at a population of ~150,000 stars in the northern constellation of Cygnus [19]. The goal of *Kepler* was to perform the first exoplanet census, to reveal the true abundance of planets around other stars. In this, the spacecraft was successful beyond anyone's wildest imaginings. To date, observations taken during the main phase of *Kepler*'s mission have led to the discovery of 2,341 confirmed transiting planets, and a further 2,155 candidate planets (e.g. [20][21]), whilst those taken in the more recent K2 program ([22]) have yielded a further 197 confirmed and 425 candidate planets [23].

One of the chief legacies of the *Kepler* mission is that we now know that a majority of stars host small planets with orbital periods less than ~200 days. The depth of the transit yields a planetary radius measurement, though this is critically dependent on the accuracy and precision of the physical parameters of the host star [24],[25],[26]. The planetary radius combined with an estimate of the planet's mass (from radial velocity) then delivers the bulk density, which in turn informs models of the composition of these planets (e.g. [27][28][29]). Thanks to the thousands of confirmed *Kepler* planets, we are getting a sense of the broad range of exoplanetary compositions, from densities exceeding that of solid iron (e.g. [30][31]) to bloated worlds evaporating from the intensity of their host star's light (e.g. [32][33]). The detailed properties of planetary systems are *far more diverse* than could have been imagined less than ten years ago.

The *Kepler* mission has shown that small planets are common, and that "super-Earths" (planets of 2-10 Earth masses) are the most common of all ([34][35]). But such planets are completely absent from our own Solar system, which is clearly not representative of the general population of planetary system architectures. If we want to know what other planetary systems *do* look like, we must substantially improve the observational database of super-Earth planets with measured masses and radii.

The next step in this journey will come with the launch of NASA's next generation exoplanet survey spacecraft, the Transiting Exoplanet Survey Satellite, TESS [36]. Where *Kepler*'s primary mission focussed on a single small region of the night sky, TESS will instead perform an all-sky survey. Launched on 2018 April 18, TESS' primary mission is intended to last for two years, during which time the satellite will survey approximately 200,000 of the brightest stars in the sky with a two-minute observing cadence to search for evidence of transiting companions. In addition, the spacecraft will return data on an estimated 20 million stars from the full frame images, with a cadence of 30 minutes.

TESS features four wide-angle cameras, with field of view approximately 24° x 24°. Taken together, these cameras yield a rectangular field of view of approximately 96° x 24°. The spacecraft will be oriented such that one of the four cameras will be centred on one of the ecliptic poles, with the others pointing progressively closer to the ecliptic. That 24° wide stripe of the celestial sphere will be monitored continuously for a period of approximately 27 days, then the spacecraft will rotate in the plane of the ecliptic to monitor the adjacent strip of the sky. In this manner, TESS will survey the southern ecliptic hemisphere of the sky in its first year of operation, before moving on to observe the northern hemisphere sky for the second of its mission. Stars within a few degrees of the ecliptic will not be observed at all in the primary mission.

Most stars will be observed for a period of just 27 days using this technique, which means that TESS will be strongly biased towards the detection of planets moving on short-period orbits. However, the fields closer to the ecliptic poles will overlap, such that there will be a region of the sky for which stars will be monitored for a full year. Where *Kepler* focussed on faint stars, those observed by TESS will be bright enough to be readily observed from the ground, which will facilitate ground-based follow-up of the thousands of planets that the spacecraft is expected to yield – helping to solve the current paucity of known super-Earths with well determined masses and radii.

Figure 1, adapted from Figure 9 of [36], shows the current severe shortage of readily characterisable planets (left panel) and the expected contribution from TESS (right panel).

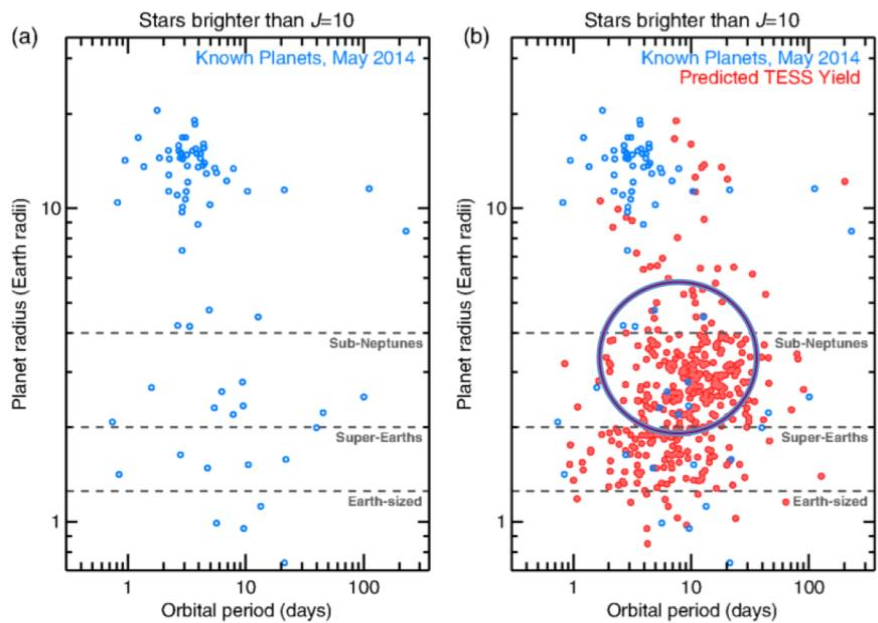

*Figure 1: Whilst Kepler has revealed that planets smaller than Neptune are common in the cosmos, very few of them have to date been fully characterised (left). This is the result of the vast majority of Kepler target stars being too faint to be followed up with radial velocity observations from the ground, coupled to the lack of ground-based facilities dedicated to such work. The situation will be remedied once NASA's TESS mission is complete (right), with TESS expected to discover thousands of such planets orbiting stars bright enough for ground based follow-up and characterisation. Figured adapted from Figure 9 of [36].*

The stars targeted by TESS will be, on average, about 100 times brighter than *Kepler* stars, greatly improving the prospects for detailed follow-up. At present, only a few tens of planets that cover the critical and poorly understood transition between Earth-like and Neptune-like compositions have sufficiently precise mass measurements to distinguish their compositions. Improved spectroscopic stellar parameters (e.g. [37][38]) have helped to identify a gap in the radius distribution (e.g. [39][40]). For improved mass measurements, more planets are needed around brighter host stars – TESS will deliver many hundreds of such planets [41].

The large number of *Kepler* candidate planets that remain to be characterised reveals a persistent challenge for exoplanetary scientists. Whilst transit surveys can yield large numbers of potential planets, they require significant resources to be dedicated to follow-up work to convert these 'candidates' into 'confirmed' worlds. Furthermore, truly understanding the detailed properties of these planets requires a prodigious observational effort. Indeed, only about 1% of *Kepler*'s planets have mass measurements. Quite simply, we are in a situation where we have too many planets, and too few telescopes to confirm them.

In order to address this for planets in the northern sky, an international consortium obtained funding to build the MINERVA facility – a dedicated exoplanet confirmation and characterisation facility, located at Mt. Hopkins, in the Arizona desert. MINERVA features four 70cm telescopes that feed light directly to a high-resolution spectrograph [42]. The primary mission of the Northern MINERVA observatory is to perform a high-cadence radial velocity search for small planets orbiting nearby bright stars. As a dedicated facility, MINERVA can observe potential planetary candidates all night, every night.

The flexibility of having multiple dedicated telescopes allows the MINERVA array to yield a significant amount of ancillary science, as has been revealed by the first results published by the consortium using data obtained during commissioning of the array. From the detection of a crumbling, disintegrating world orbiting a white dwarf star (e.g. [43][44]) to the observation of a lengthy dimming even from a young star obscured by the disk of its companion ([45]), the benefits of researchers having access to a dedicated facility are being amply demonstrated.

Such follow-up work has additional benefits, beyond merely confirming those planets we suspect to be there. Among planets discovered using the radial velocity technique, there exist a subset that appear to move on highly eccentric orbits (e.g. [46][47][48]). In recent years, such single eccentric planets have come under significant scrutiny, with a growing realisation that those whose eccentricity is moderate rather than extreme may instead represent multiple planet systems, featuring planets on near-circular orbits (e.g. [49][50][51]). Follow-up observations of such systems are therefore of vital importance to help support or disprove the single-planet hypothesis. The deeper investigation of these more unusual exoplanets is doubly valuable, since researchers attempting to understand planet occurrence and formation must attempt to match their results with the observed database – and if that database is polluted with unusual proposed systems that either do not exist (e.g. [52][53][54]) or move on dramatically different orbits to those published (e.g. [55][56][57]), this will lead to problems in the field's ongoing growth and development.

For this reason, we are currently in the process of constructing the MINERVA-Australis facility at the University of Southern Queensland's Mount Kent Observatory, building on the template and groundwork laid by the northern hemisphere MINERVA team. MINERVA-Australis is primarily intended to support the work of NASA's TESS mission ([36]). TESS will discover thousands of new candidate exoplanets, scattered across both the northern and southern hemisphere skies, and it is vital importance to ensure that sufficient follow-up capacity is available to ensure the maximum return on this exceptional scientific resource.

# The MINERVA-Australis facility

MINERVA-Australis comprises six 0.7m PlaneWave CDK-700 telescopes, all feeding a single Kiwispec high-resolution spectrograph. The MINERVA model uses multiple small telescopes to achieve the same light-collecting power of a single larger telescope at a fraction of the cost and build time. Figure 2, reprised from Figure 1 of [42], shows the comparison in cost as a function of effective aperture for MINERVA's telescopes versus custom-built monolithic telescopes [58].

The MINERVA-Australis telescopes will simultaneously feed a single spectrograph via fibre optic cables. The fibres are aligned in the cross-dispersion direction of the spectrometer to form six individual echelle traces. Fibre-feeding the high resolution (R~80,000) spectrograph provides a stable instrumental profile through fibre scrambling ([59][60]), and allows it to be bench-mounted and housed in an insulated, environmentally controlled enclosure. The spectrograph will be housed in a purpose-built class 100,000 cleanroom, with the critical components inside a vacuum chamber and thermally stabilised to ±0.01 K. The spectrograph point-spread function is calibrated using an iodine absorption cell in the light path, a well-established technique (e.g. [61][62]).

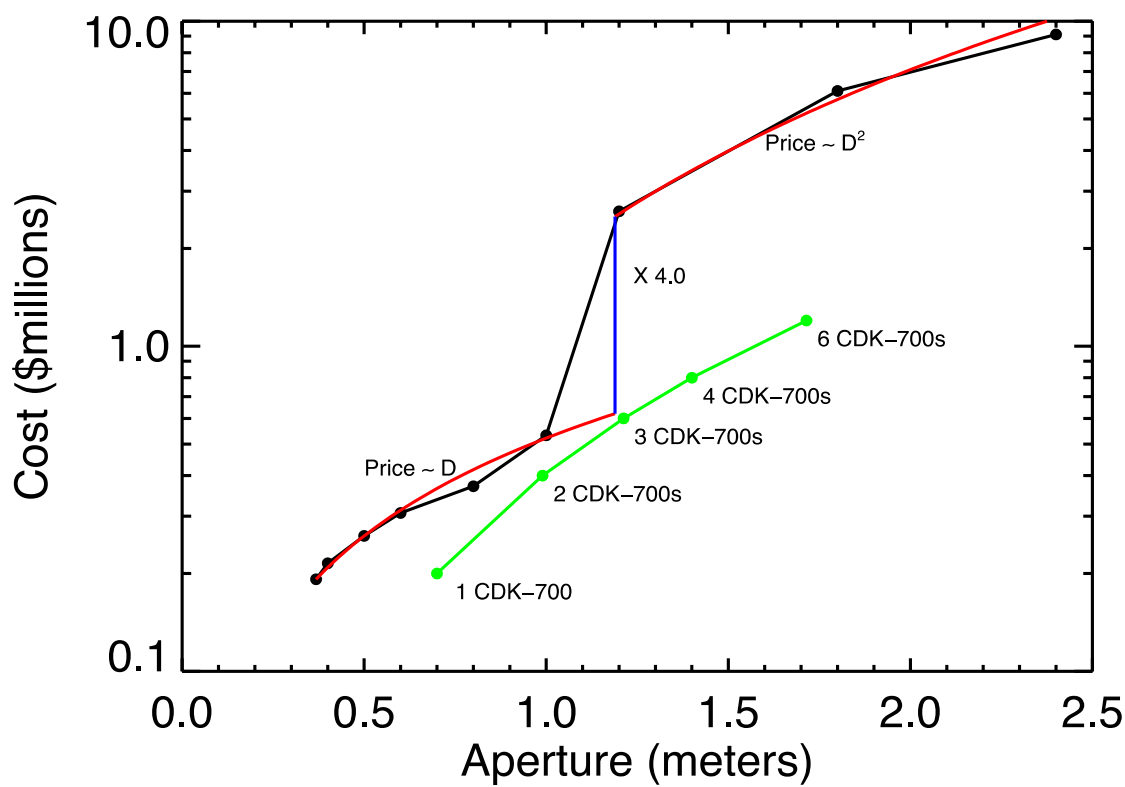

*Figure 2: The benefit of purchasing telescopes 'off-the-shelf', compared with custom-built instruments. By taking advantage of the relatively low cost of PlaneWave Instrument's CDK-700 telescopes, MINERVA-Australis will have an effective collecting area equivalent to a single telescope with a 1.7m diameter mirror, at a fraction of the cost.*

Based on the performance of the Northern MINERVA facility, which uses identical hardware and software, and scaling by photon statistics, we benchmark a radial velocity precision of 3 m/s in 20 minutes on a V=10 star, using all six telescopes. To estimate the number of TESS planets for which MINERVA-Australis can obtain mass estimates, we used the TESS yield simulations of [41], taking as input planet candidates the results given in their Table 6 (RA/dec, V magnitude,

RV amplitude *K*). Scaling from our performance metric above, assuming 60% usable time, and only considering Southern hemisphere targets (Declination < 0), we estimate MINERVA-Australis will be able to validate ~200 planets in the first three years. A planet is considered validated if we obtain 20 velocity measurements with a precision of 3 m/s. A second estimate, for which we instead required that $K_{planet} \times \sqrt{N}/\sigma > 7.5$ ([63]), yielded a similar number of MINERVA-Australis mass measurements, again including most TESS targets with V<10.5.

Being dedicated to TESS follow-up, MINERVA-Australis holds a unique position among the Southern hemisphere's precision radial velocity instruments. The high cadence afforded by MINERVA-Australis will permit highly precise mass measurements (often requiring more than 100 radial velocity observations), and hence more detailed characterisation of planetary compositions. Continued monitoring of TESS planetary systems will almost certainly reveal a population of outer, non-transiting planets, as has already been demonstrated in the follow-up of some *Kepler* systems (e.g. [64][65]). Single-transit events can also be monitored and their planets recovered thanks to the flexible scheduling of MINERVA-Australis.

As at 2018 April, the site works are complete, and the spectrograph will be delivered in May, with commissioning in May and June using Telescope 1. We anticipate operations with at least four telescopes by Q4 2018 in time for the TESS data release scheduled for December 2018.

## Ancillary Science and the Future

In the coming years, our primary focus will be to follow up on the plethora of new worlds that will be discovered by the TESS mission. As a dedicated, guaranteed time facility, however, MINERVA-Australis will have the flexibility to perform ancillary observations.

During the commissioning phase, we intend to start a program of radial-velocity follow-up of targets that were previously observed as part of the Anglo-Australian Planet Search (AAPS; e.g. [66][67]). That survey, which was undertaken using the 3.9m Anglo-Australian Telescope between 1998 and 2014, gathered radial velocity observations of more than 240 southern stars. In its final years, AAPS began to discover true Jupiter-analogues – planets of mass comparable to the Solar system's giant planets, moving on orbits with periods of a decade or more (e.g. [68][69][70]). The discovery of such planets is particularly challenging – the long orbital periods require an incredibly lengthy observational dataset, and as such, the population of such planets remains poorly studied. We will use the commissioning period of MINERVA-Australis to begin the AAPS legacy survey – targeting those stars in the AAPS sample for which the data hints at the potential presence of long-term radial velocity trends. By continuing the AAPS survey in this way, we hope to expand the population of known Jupiter analogues, and thereby improve our understanding of the degree to which our own Solar system is unusual or unique.

In recent years, observations of Solar system small bodies occulting background stars have yielded a wealth of vital information. Such occultations significantly improve the precision with which we know the orbits of Solar system bodies (e.g. [71]), as well as allowing the precise determination of the size and shape of those objects (e.g. [72][73]). Occultation observations have also led to the discovery of ring systems around several of the Solar system's minor bodies (e.g. [74][75][76]), which has greatly increased our understanding of those objects formation and evolution (e.g. [77][78][79]). In the coming years, such occultation observations will become ever more important in our understanding of objects throughout the Solar system, and will be facilitated by the release of data obtained by the GAIA spacecraft (e.g. [80][81]), which will significantly improve the precision with which we know the locations of the stars in the night sky. That improved precision will make it far easier to predict the timing of future occultation events, as well as helping to localise the regions on Earth where they might be observed. To capitalise on this, we have applied for funding to purchase two additional cameras for the MINERVA-Australis array, with the goal of being able to participate in target-of-opportunity observations of high-priority occultation events.

## Summary and Conclusion

We are privileged to be in the first generation of humans to know that many of the points of light dusting our night sky are host to orbiting worlds, some of which may be like our Earth. The NASA TESS mission will deliver thousands of planets that can only be fully understood with ground-based follow-up observations. MINERVA-Australis, located at Mount Kent Observatory in southeast Queensland, is a new Australian facility wholly dedicated to unlocking the secrets of the diversity of these new worlds. In the coming decade, MINERVA-Australis will be instrumental in establishing Australia's leadership in exoplanetary and Solar system science.

## Acknowledgements

MINERVA-Australis is a partnership among the University of Southern Queensland, UNSW Sydney, MIT, George Mason University, University of California Riverside, University of Louisville, University of Texas, University of Florida, and Nanjing University, and is funded in part by Australian Research Council LIEF grant LE160100001, ARC Discovery Grant DP180100972, and the Mount Cuba Astronomical Foundation.

## References

1. Aumann, H. H., Beichman, C. A., Gillett, F. C. et al., "Discovery of a shell around Alpha Lyrae", 1984, *The Astrophysical Journal*, 278, 23
2. Neugebauer, G., Soifer, B. T., Beichman, C. A. et al., "Early results from the Infrared Astronomical Satellite", 1984, *Science*, 224, 14
3. Gillett, F. C., "IRAS observations of cool excess around main sequence stars", 1986, *Light on dark matter; Proceedings of the First Infra-Red Astronomical Satellite Conference*, Noordwijk, Netherlands, June 10-14, 1985; Dordrecht, D. Reidel Publishing Co., p. 61 - 69
4. Wolszczan, A. & Frail, D. A., "A planetary system around the millisecond pulsar PSR1257 +12", 1992, *Nature*, 355, 145
5. Mayor, M. & Queloz, D., "A Jupiter-mass companion to a solar-type star", 1995, *Nature*, 378, 355
6. Butler, R. P. & Marcy, G. W., "A Planet Orbiting 47 Ursae Majoris", 1996, *The Astrophysical Journal*, 464, 153
7. Marcy, G. W. & Butler, R. P., "A Planetary Companion to 70 Virginis", 1996, *The Astrophysical Journal*, 464, 147
8. Latham, D. W., Stefanik, R. P., Mazeh, T. et al., "The unseen companion of HD114762 – A probable brown dwarf", 1989, *Nature*, 339, 38
9. Campbell, B., Walker, G. A. H. & Yang, S., "A search for substellar companions to solar-type stars", *The Astrophysical Journal*, 331, 902
10. Santos, N. C., Bouchy, F., Mayor, M. et al., "The HARPS survey for southern extra-solar planets. II. A 14 Earth-masses exoplanet around μ Arae", 2004, *Astronomy and Astrophysics*, 426, 19
11. Butler, R. P., Wright, J. T., Marcy, G. W., et al., "Catalog of Nearby Exoplanets", 2006, *The Astrophysical Journal*, 646, 505

12. Robertson, P., Endl, M., Cochran, W. D. et al., "The McDonald Observatory Planet Search: New Long-period Giant Planets and Two Interacting Jupiters in the HD 155358 System", 2012, *The Astrophysical Journal*, 749, 39
13. Marcy, G. W., Butler, R. P., Fischer, D. A. et al., "A Planet at 5 AU around 55 Cancri", 2002, *The Astrophysical Journal*, 581, 1375
14. Boisse, I., Pepe, F., Perrier, C. et al., "The SOPHIE search for northern extrasolar planets. V. Follow-up of ELODIE candidates: Jupiter-analogs around Sun-like stars", 2012, *Astronomy and Astrophysics*, 545, 55
15. Robertson, P., Horner, J., Wittenmyer, R. A. et al., "A Second Giant Planet in 3:2 Mean-motion Resonance in the HD 204313 System", 2012, *The Astrophysical Journal*, 754, 50
16. Konacki, M., Torres, G., Jha, S. & Sasselov, D. D., "An extrasolar planet that transits the disk of its parent star", 2003, *Nature*, 421, 6922, 507
17. Alonso, R., Brown, T. M., Torres, G. et al., "TrES-1: The Transiting Planet of a Bright K0 V Star", 2004, *The Astrophysical Journal*, 613, 2, L153
18. Bakos, G. Á., Noyes, R. W., Kovács, G. et al., "HAT-P-1b: A Large-Radius, Low-Density Exoplanet Transiting One Member of a Stellar Binary", 2007, *The Astrophysical Journal*, 656, 552
19. Koch, D. G., Borucki, W. J., Basri, G. et al., "Kepler Mission Design, Realized Photometric Performance, and Early Science", 2010, *The Astrophysical Journal Letters*, 713, L79
20. Borucki, W. J., Koch, D., Basri, G. et al., "Kepler Planet-Detection Mission: Introduction and First Results", 2010, *Science*, 327, 5968, 977
21. Coughlin, J. L., Mullally, F., Thompson, S. E., et al. "Planetary Candidates Observed by Kepler. VII. The First Fully Uniform Catalog Based on the Entire 48-month Data Set (Q1-Q17 DR24), 2016, *The Astrophysical Journal Supplement Series,* 224, 12
22. Howell, S. B., Sobeck, C., Hass, M. et al., "The K2 Mission: Characterization and Early Results", 2014, *Publications of the Astronomical Society of the Pacific*, 126, 938, 398
23. The latest statistics on planetary discoveries made by the Kepler and K2 missions can be found at https://www.nasa.gov/kepler/discoveries. The values in the text were obtained from that website on 1st February 2018.
24. Petigura, E.A., Howard, A.W., Marcy, G.W., et al. 2017, *The Astronomical Journal*, 154, 107
25. Johnson, J.A., Petigura, E.A., Fulton, B.J., et al. 2017, *The Astronomical Journal*, 154, 108
26. Wittenmyer, R.A., Sharma, S., Stello, D., et al. 2018, *The Astronomical Journal*, 155, 84
27. Seager, S., Kuchner, M., Hier-Majumder, C. A. & Militzer, B., "Mass-Radius Relationships for Solid Exoplanets", 2007, *The Astrophysical Journal,* 669, 1279
28. Rogers, L. A. & Seager, S., "A Framework for Quantifying the Degeneracies of Exoplanet Interior Compositions", 2010, *The Astrophysical Journal,* 712, 974
29. Wolfgang, A. & Lopez, E., "How Rocky Are They? The Composition Distribution of Kepler's Sub-Neptune Planet Candidates within 0.15 AU", 2015, *The Astrophysical Journal,* 806, 183
30. Batalha, N. M., Borucki, W. J., Bryson, S. T., et al., "Kepler's First Rocky Planet: Kepler-10b", 2011, *The Astrophysical Journal,* 729, 27
31. Wu, Y. & Lithwick, Y., "Density and Eccentricity of Kepler Planets", 2013, *The Astrophysical Journal,* 772, 74
32. Rappaport, S., Levine, A., Chiang, E. et al., "Possible Disintegrating Short-period Super-Mercury Orbiting KIC 12557548", 2012, *The Astrophysical Journal,* 752, 1

33. van Werkhoven, T. I. M., Brogi, M., Snellen, I. A. G. & Keller, C. U., "Analysis and interpretation of 15 quarters of Kepler data of the disintegrating planet KIC 12557548 b", 2014, *Astronomy and Astrophysics*, 561, 3
34. Fressin, F., Torres, G., Charbonneau, D. et al., "The False Positive Rate of Kepler and the Occurrence of Planets", 2013, *The Astrophysical Journal,* 766, 81
35. Howard, A. W., Marcy G. W., Bryson, S. T. et al., "Planet Occurrence within 0.25 AU of Solar-type Stars from Kepler", 2017, *The Astrophysical Journal Supplement,* 201, 15
36. Ricker, G. R., Winn, J. N., Vanderspek, R. et al., "Transiting Exoplanet Survey Satellite (TESS)", 2015, *Journal of Astronomical Telescopes, Instruments, and Systems,* 1, 014003
37. Johnson, J. A., Petigura, E. A., Fulton, B. J. et al., "The California-Kepler Survey. II. Precise Physical Properties of 2025 Kepler Planets and Their Host Stars", 2017, *The Astronomical Journal,* 154, 108
38. Wittenmyer, R. A., Sharma, S., Stello, D., et al., "The K2-HERMES Survey. I. Planet Candidate Properties from K2 Campaigns 1-3", 2018, *The Astronomical Journal, in press*
39. Fulton, B. J., Petigura, E. A., Howard, A. W. et al., "The California-Kepler Survey. III. A Gap in the Radius Distribution of Small Planets", 2017, *The Astronomical Journal*, 154, 109
40. Owen, J. E. & Wu, Y., "The Evaporation Valley in the Kepler Planets", 2017, *The Astrophysical Journal*, 847, 29
41. Sullivan, P. W., Winn, J. N., Berta-Thompson, Z. K. et al., "The Transiting Exoplanet Survey Satellite: Simulations of Planet Detections and Astrophysical False Positives", 2015, *The Astrophysical Journal*, 809, 77
42. Swift, J. J., Bottom, M., Johnson, J. A., et al., "Miniature Exoplanet Radial Velocity Array (MINERVA) I. Design, Commissioning, and First Science Results", 2015, *Journal of Astronomical Telescopes, Instruments, and Systems*, 1, 027002
43. Vanderburg, A., Johnson, J. A., Rappaport, S., et al., "A disintegrating minor planet transiting a white dwarf", 2015, *Nature*, 526, 546
44. Croll, B., Dalba, P. A., Vanderburg, A., "Multiwavelength Transit Observations of the Candidate Disintegrating Planetesimals Orbiting WD 1145+017", 2017, *The Astrophysical Journal*, 836, 82
45. Rodriguez, J. E., Zhou, G., Cargile, P. A. et al., "The Mysterious Dimmings of the T Tauri Star V1334 Tau", 2017, *The Astrophysical Journal*, 836, 209
46. Naef, D., Mayor, M., Lo Curto, G. et al., "The HARPS search for southern extrasolar planets. XXIII. 8 planetary companions to low-activity solar-type stars", 2010, *Astronomy and Astrophysics,* 523, 15
47. Kane, S. R., Wittenmyer, R. A., Hinkel, N. R. et al., "Evidence for Reflected Light from the Most Eccentric Exoplanet Known", 2016, *The Astrophysical Journal*, 821, 65
48. Wittenmyer, R. A., Jones, M. I., Horner, J. et al., "The Pan-Pacific Planet Search. VII. The Most Eccentric Planet Orbiting a Giant Star", 2017, *The Astronomical Journal*, 154, 274
49. Anglada-Escudé, G., López-Morales, M. & Chambers, J. E., "How Eccentric Orbital Solutions Can Hide Planetary Systems in 2:1 Resonant Orbits", 2010, *The Astrophysical Journal*, 709, 168
50. Wittenmyer, R. A., Horner, J., Tuomi, M. et al., "The Anglo-Australian Planet Search. XXII. Two New Multi-planet Systems", 2012, *The Astrophysical Journal*, 169, 12
51. Wittenmyer, R. A., Wang, S., Horner, J. et al., "Forever Alone? Testing Single Eccentric Planetary Systems for Multiple Companions", 2013, *The Astrophysical Journal: Supplement*, 208, 2

52. Horner, J., Marshall, J. P., Wittenmyer, R. A. & Tinney, C. G., "A dynamical analysis of the proposed HU Aquarii planetary system", 2011, *Monthly Notices of the Royal Astronomical Society*, 416, 11
53. Wittenmyer, R. A., Horner, J. & Marshall, J. P., "On the dynamical stability of the proposed planetary system orbiting NSVS 14256825", 2013, *Monthly Notices of the Royal Astronomical Society,* 431, 2150
54. Horner, J., Wittenmyer, R. A., Hinse, T. C. et al., "A detailed dynamical investigation of the proposed QS Virginis planetary system", 2013, *Monthly Notices of the Royal Astronomical Society*, 435, 2033
55. Wittenmyer, R. A., Horner, J. & Tinney, C. G., "Resonances Required: Dynamical Analysis of the 24 Sex and HD 200964 Planetary Systems", 2012, *The Astrophysical Journal*, 761, 165
56. Wittenmyer, R. A., Johnson, J. A., Butler, R. P. et al., "The Pan-Pacific Planet Search. IV. Two Super-Jupiters in a 3:5 Resonance Orbiting the Giant Star HD 33844", 2016, *The Astrophysical Journal*, 818, 35
57. Horner, J., Wittenmyer, R. A., Wright, D. J. et al., "The HD 181433 Planetary System: Dynamics and a new Planetary Solution", 2018, *The Astronomical Journal, in press*
58. Survey conducted in 2010 and may not reflect the current state of the market.
59. Spronck, J. F. P., Schwab, C. & Fischer, D. A., "Fiber-stabilized PSF for sub-m/s Doppler precision at Lick Observatory", 2010, *Proceedings of the SPIE*, 7735, 77350
60. Perruchot, S., Bouchy, F., Chazelas, B. et al., "Higher-precision radial velocity measurements with the SOPHIE spectrograph using octagonal-section fibers", 2011, *Proceedings of the SPIE*, 8151, 815115
61. Marcy, G. W. & Butler, R. P., "Precision radial velocities with an iodine absorption cell", 1992, *Publications of the Astronomical Society of the Pacific*, 104, 270
62. Buttler, R. P., Marcy, G. W., Williams, E. et al., "Attaining Doppler Precision of 3 ms$^{-1}$", 1996, *Publications of the Astronomical Society of the Pacific*, 108, 500
63. Dumusque, X., Borsa, F., Damasso, M. et al., "Radial-velocity fitting challenge. II. First results of the analysis of the data set", 2017, *Astronomy and Astrophysics*, 598, 133
64. Buchhave, L. A., Dressing, C. D., Dumusque, X. et al., "A 1.9 Earth Radius Rocky Planet and the Discovery of a Non-transiting Planet in the Kepler-20 System", 2016, *The Astronomical Journal*, 152, 160
65. Christiansen, J. L., Vanderburg, A., Burt, J. et al., "Three's Company: An Additional Non-transiting Super-Earth in the Bright HD 3167 System, and Masses for All Three Planets", 2017, *The Astronomical Journal*, 154, 122
66. Tinney, C. G., Butler, R. P., Marcy, G. W. et al., "Two Extrasolar Planets from the Anglo-Australian Planet Search", 2002, *The Astrophysical Journal*, 571, 528
67. Tinney, C. G., Wittenmyer, R. A., Butler, R. P. et al., "The Anglo-Australian Planet Search. XXI. A Gas-giant Planet in a One Year Orbit and the Habitability of Gas-giant Satellites", 2011, *The Astrophysical Journal*, 732, 31
68. Wittenmyer, R. A., Horner, J., Tinney, C. G. et al., "The Anglo-Australian Planet Search. XXIII. Two New Jupiter Analogs", 2014, *The Astrophysical Journal*, 783, 103
69. Wittenmyer, R. A., Butler, R. P., Tinney, C. G. et al., "The Anglo-Australian Planet Search XXIV: The Frequency of Jupiter Analogs", 2016, *The Astrophysical Journal*, 819, 28
70. Wittenmyer, R. A., Horner, J., Mengel, M. W. et al., "The Anglo-Australian Planet Search. XXV. A Candidate Massive Saturn Analog Orbiting HD 30177", 2017, *The Astronomical Journal*, 153, 167
71. Young, E., "Mission Support of the New Horizons 2014 MU69 Encounter via Stellar Occultations", 2016, *SOFIA Proposal, Cycle 5, ID. 05_0168*
72. Sicardy, B., Ortiz, J. L., Assafin, M. et al., "A Pluto-like radius and a high albedo for the dwarf planet Eris from an occultation", 2011, *Nature*, 478, 493

73. Ortiz, J. L., Sicardy, B., Braga-Ribas, F. et al., "Albedo and atmospheric constraints of dwarf planet Makemake from a stellar occultation", 2012, *Nature*, 491, 566
74. Braga-Ribas, F., Sicardy, B., Ortiz, J. L. et al., "A ring system detected around the Centaur (10199) Chariklo", 2014, *Nature*, 508, 72
75. Ortiz, J. L., Duffard, R., Pinilla-Alonso, N. et al., "Possible ring material around centaur (2060) Chiron", 2015, *Astronomy and Astrophysics*, 576, 18
76. Ortiz, J. L., Santos-Sanz, P., Sicardy, B. et al., "The size, shape, density and ring of the dwarf planet Haumea from a stellar occultation", 2017, *Nature*, 550, 219
77. Wood, J., Horner, J., Hinse, T. C. & Marsden, S. C., "The Dynamical History of Chariklo and Its Rings", 2017, *The Astronomical Journal*, 153, 245
78. Wood, J., Horner, J., Hinse, T. C. & Marsden, S. C., "The Dynamical History of 2060 Chiron and Its Proposed Ring System", 2018, *The Astronomical Journal*, 155, 2
79. Vilenius, E., Stansberry, J., Müller, T. et al., ""TNOs are Cool": A survey of the trans-Neptunian region XIII. Size/Albedo characterization of the Haumea family observed with *Herschel* and *Spitzer*"", 2018, *Astronomy and Astrophysics*, *in press*
80. Tanga, P. & Delbo, M., "Asteroid occultations today and tomorrow: toward the GAIA era", 2007, *Astronomy and Astrophysics,* 474, 3
81. Sporto, F., Tanga, P., Bouquillon, S., Desmar, J. et al., "Ground-based astrometry calibrated by Gaia DR1: new perspectives in asteroid orbit determination", 2017, *Astronomy & Astrophysics,* 607, 21